# Using Mathcasts to Facilitate Student Comprehension of Physical Applications of Math Concepts

Colleen Lanz Countryman and Hong Wang, North Carolina State University, Raleigh, NC

Many Physics Education Researchers have discussed the positive correlation between students' incoming mathematics skills and performance in their physics classes[1,2]. Thus, in order to strengthen their performance gains in their physics courses, professors strive to bolster students' mathematics abilities but in-class time constraints force instructors to focus their lesson plans on physics concepts rather than renewing mathematics material.

For the sake of preserving in-class time, many instructors are turning to lecture delivery outside of class by way of flipped classrooms and MOOCs (Massive Online Open Courses). Videos can strengthen problem-solving skill sets[3] and some PER researchers have begun moving worked problems and class lectures to YouTube[4,5]. YouTube videos even have a relatively long history of being used for in-class discussions[6].

We aimed to provide students with videos that directly connected mathematics concepts from previous classes (such as dot/cross products, limits of integration, solving systems of equations, etc.) to E&M concepts in their current class (such as work/torque, electric fields from continuous objects, Kirchoff's Laws, etc.)[7].

## Lessons Regarding Video Production

During the production of our series of six videos throughout the semester, we took into account several lessons learned from an empirical study on videos intended for use in MOOCs[8]. We tried to keep the videos short and personal. We did not make use of PowerPoint slides and only displayed typewritten text with the problem statement. Where appropriate (such as during discussions of the right hand rule), we interspersed a "talking head" and demonstrations using physical objects. For each of the videos, the instructor spoke fairly quickly and with enthusiasm to sustain students' interest.

We chose to model our videos after the Khan Academy[9] tutorials primarily due to their increasing popularity and because they allow the instructor to "situate themselves 'on the same level' as the student rather than talking *at* the student in 'lecturer mode.'"[8] In addition, it allows for hand-sketching and handwriting which have both been shown to increase student engagement over computer-rendered graphics and text[10,11].

## Video Creation/YouTube Analytics

We opted for one of the least expensive methods for video creation. The only purchase required was for a Wacom Bamboo tablet, which retails for approximately

$80. This allows the instructor to use a stylus as opposed to a mouse to transfer their sketches and writing on screen. While a variety of screen-capture software and drawing programs exists[12], we chose two free Mac-supported programs, SketchBook Express and Quicktime Player. The instructor simply prepares the "blackboard" in the drawing program and chooses "New Screen Recording…" in the File menu of Quicktime in order to begin recording a selected region of the screen.

YouTube is a free video hosting website that worked well for our purposes. Privacy options within YouTube allow the instructor to make the videos accessible to the public or to only those with the link. YouTube also provides tools for analytics which allow the instructor to monitor views, demographics of viewers, playback locations, retention, comments, etc.

## Results from our study

Pre- and post-video assignments were administered to students. Due to in-class time constraints, some assignments were administered in class on paper while others were administered through an online instructional website, WebAssign. When the latter was used, students' first responses to the questions were used in order to mimic performance on an in-class assignment. With the exception of the first video, students' performance increased from pre- to post-video assignments. Results from all six of the videos are illustrated in figure 1.

In addition, five of the post-video assignments also included a survey which aimed to assess students' thoughts on the effectiveness of the video. Figure 2 includes the results of this survey, and demonstrates how students generally perceived the videos as more helpful as the semester went on.

We also allowed students an opportunity to provide us with additional open-ended feedback on these surveys. These comments allowed us to alter the format of upcoming videos as needed. We primarily observed students being grateful for the videos as they allowed them to regain skills they had lost or reaffirm their thoughts regarding specific concepts. One student made a special note of usefulness of the demonstrations, indicating that they were "very useful" and provided a "good change of pace."

## Limitations of the Study and Areas for Future Work

While these YouTube videos appear to have encouraged students to draw direct connections from their math classes to their physics classes, we did make note of several drawbacks of using this format. First, there is no easy way to individualize instruction for specific students. In addition, if the videos are made public, there is no clear way to differentiate students' data (in aggregate) from the general public's data.

We attempted to minimize cheating and increase the number of students that actually watched the video by requiring that they enter into a WebAssignment a password (which is displayed randomly in the video for a few seconds with YouTube's annotation feature). While we expect that this cuts back on the number of students who falsely claim to have watched the video, there is still a chance that students will receive the password from another classmate and enter it into the WebAssignment without actually having watched the video.

Looking forward, we anticipate that these videos will be a required part of the traditional E&M coursework. Due to the feedback from students that we received, we would like to attempt to make more videos and fine tune the videos that we've already created.

## Acknowledgments

We would like to thank Dr. Robert Beichner, Dr. M. A. Paesler, Dr. Lili Cui and Dr. David G. Haase for their helpful comments. This study was possible due to support from the Scholarship for Teaching and Learning at North Carolina State University.

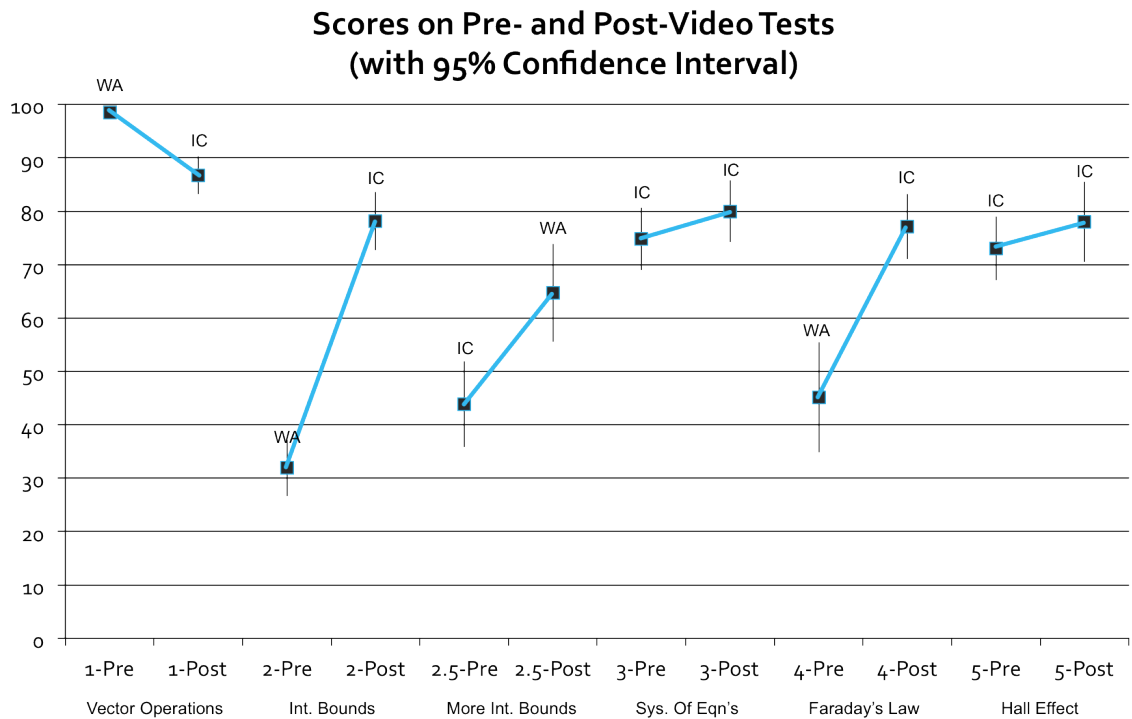

**Figure 1:** Data points are labeled "WA" and "IC" for assignments administered on the online instructional system, WebAssign, and in-class on paper, respectively. Error bars represent a 95% confidence interval. Blue lines connect the pre- and post-video data points for each video.

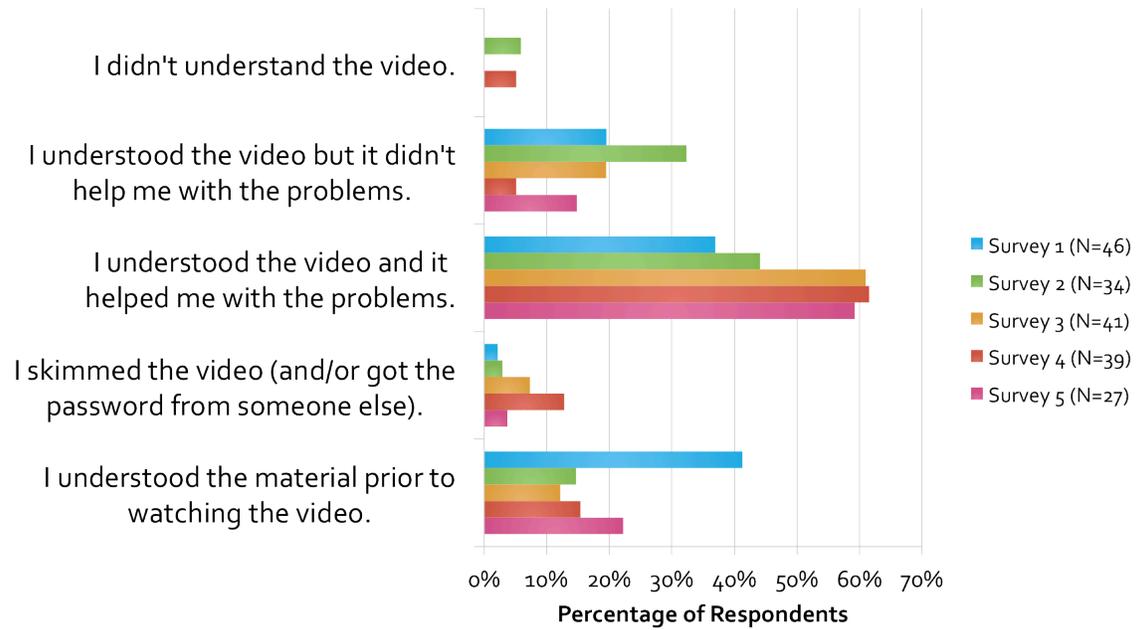

**Figure 2:** The results of an anonymous survey attached to five of the post-video assignments. The last three videos were deemed more understandable and helpful than the first two.